\documentstyle[11pt,
epsfig]{article}

\begin{document}

\relax

\def\be{\begin{equation}}
\def\ee{\end{equation}}
\def\bs{\begin{subequations}}
\def\es{\end{subequations}}
\def\calm{{\cal M}}
\def\lx{\lambda}
\def\ex{\epsilon}
\def\Lx{\Lambda}

\newcommand{\mut}{\tilde{\mu}}
\newcommand{\vrm}{{\rm v}}
\newcommand{\tl}{\tilde t}
\newcommand{\ttt}{\tilde T}
\newcommand{\rhot}{\tilde \rho}
\newcommand{\ptt}{\tilde p}
\newcommand{\drho}{\delta \rho}
\newcommand{\drhot}{\delta {\tilde \rho}}
\newcommand{\dchi}{\delta \chi}
\newcommand{\A}{A}
\newcommand{\B}{B}
\newcommand{\mmu}{\mu}
\newcommand{\mnu}{\nu}
\newcommand{\ii}{i}
\newcommand{\jj}{j}
\newcommand{\jl}{[}
\newcommand{\jr}{]}
\newcommand{\ml}{\sharp}
\newcommand{\mr}{\sharp}

\newcommand{\da}{\dot{a}}
\newcommand{\db}{\dot{b}}
\newcommand{\dn}{\dot{n}}
\newcommand{\dda}{\ddot{a}}
\newcommand{\ddb}{\ddot{b}}
\newcommand{\ddn}{\ddot{n}}
\newcommand{\pa}{a^{\prime}}
\newcommand{\pn}{n^{\prime}}
\newcommand{\ppa}{a^{\prime \prime}}
\newcommand{\ppb}{b^{\prime \prime}}
\newcommand{\ppn}{n^{\prime \prime}}
\newcommand{\fda}{\frac{\da}{a}}
\newcommand{\fdb}{\frac{\db}{b}}
\newcommand{\fdn}{\frac{\dn}{n}}
\newcommand{\fdda}{\frac{\dda}{a}}
\newcommand{\fddb}{\frac{\ddb}{b}}
\newcommand{\fddn}{\frac{\ddn}{n}}
\newcommand{\fpa}{\frac{\pa}{a}}
\newcommand{\fpb}{\frac{\pb}{b}}
\newcommand{\fpn}{\frac{\pn}{n}}
\newcommand{\fppa}{\frac{\ppa}{a}}
\newcommand{\fppb}{\frac{\ppb}{b}}
\newcommand{\fppn}{\frac{\ppn}{n}}

\newcommand{\dphi}{\delta \phi}
\newcommand{\at}{\tilde{\alpha}}
\newcommand{\pt}{\tilde{p}}
\newcommand{\rhb}{\bar{\rho}}
\newcommand{\pb}{\bar{p}}
\newcommand{\pbb}{\bar{\rm p}}
\newcommand{\rht}{\tilde{\rho}}
\newcommand{\kt}{\tilde{k}}
\newcommand{\kb}{\bar{k}}
\newcommand{\wt}{\tilde{w}}

\newcommand{\dA}{\dot{A_0}}
\newcommand{\dB}{\dot{B_0}}
\newcommand{\fdA}{\frac{\dA}{A_0}}
\newcommand{\fdB}{\frac{\dB}{B_0}}

\def\be{\begin{equation}}
\def\ee{\end{equation}}
\def\bs{\begin{subequations}}
\def\es{\end{subequations}}
\newcommand{\een}{\end{subequations}}
\newcommand{\ben}{\begin{subequations}}
\newcommand{\beq}{\begin{eqalignno}}
\newcommand{\eeq}{\end{eqalignno}}

\def \lta {\mathrel{\vcenter
     {\hbox{$<$}\nointerlineskip\hbox{$\sim$}}}}
\def \gta {\mathrel{\vcenter
     {\hbox{$>$}\nointerlineskip\hbox{$\sim$}}}}

\def\g{\gamma}
\def\mpl{M_{\rm Pl}}
\def\ms{M_{\rm s}}
\def\ls{l_{\rm s}}
\def\l{\lambda}
\def\m{\mu}
\def\n{\nu}
\def\a{\alpha}
\def\b{\beta}
\def\gs{g_{\rm s}}
\def\d{\partial}
\def\co{{\cal O}}
\def\sp{\;\;\;,\;\;\;}
\def\r{\rho}
\def\dr{\dot r}

\def\e{\epsilon}
\newcommand{\NPB}[3]{\emph{ Nucl.~Phys.} \textbf{B#1} (#2) #3}   
\newcommand{\PLB}[3]{\emph{ Phys.~Lett.} \textbf{B#1} (#2) #3}   
\newcommand{\ttbs}{\char'134}        
\newcommand\fverb{\setbox\pippobox=\hbox\bgroup\verb}
\newcommand\fverbdo{\egroup\medskip\noindent%
                        \fbox{\unhbox\pippobox}\ }
\newcommand\fverbit{\egroup\item[\fbox{\unhbox\pippobox}]}
\newbox\pippobox
\def\tr{\tilde\rho}
\def\lb{w}
\def\bbox{\nabla^2}
\def\mt{{\tilde m}}
\def\rct{{\tilde r}_c}

\def \lta {\mathrel{\vcenter
     {\hbox{$<$}\nointerlineskip\hbox{$\sim$}}}}
\def \gta {\mathrel{\vcenter
     {\hbox{$>$}\nointerlineskip\hbox{$\sim$}}}}

\noindent
\begin{flushright}

\end{flushright} 
\vspace{1cm}
\begin{center}
{ \Large \bf Dark Energy and Dark Matter in Galaxy Halos\\} 
\vspace{0.5cm}
{N. Tetradis} 
\\
\vspace{0.5cm}
{\it University of Athens, Department of Physics, \\
University Campus, Zographou 157 84, Athens, Greece.} 
\\
\vspace{1cm}
\abstract{
We consider the possibility that the dark matter is coupled through its
mass to a scalar field associated with the dark energy of the Universe.
In order for such a field to play a role at the present cosmological 
distances, it must be effectively massless at galactic 
length scales.
We discuss the effect of the field on the distribution of dark
matter in galaxy halos. We show that the profile of the distribution 
outside the galaxy core 
remains largely unaffected and the approximately flat rotation curves persist.
The dispersion of the dark matter velocity is enhanced by a potentially
large factor relative to the case of zero coupling between dark energy and
dark matter. The counting rates in terrestrial dark matter
detectors are similarly enhanced. Existing bounds on the properties 
of dark matter
candidates can be extended to the coupled case, by taking into
account the enhancement factor. 
\\
\vspace{1cm}
} 
\end{center}

\newpage

According to our present understanding of the matter content of the
Universe, its energy density is dominated by dark energy and dark matter. 
The usual assumption is that 
the dark energy is associated with the potential of an evolving
scalar field \cite{wetdil,peeblesold}, while the dark matter is composed
of weakly interacting massive particles. 
Usually these two sectors are assumed to interact only 
gravitationally. Attempts to resolve the coincidence problem (the 
comparable present contributions to the total energy density from the 
two components) are based on the presence of an additional interaction 
between the dark matter 
and the scalar field \cite{wetcosmon}.
The cosmological evolution now depends on the potential of the field, as well
as on the type and strength of the interaction \cite{amendola}. 
For a scalar field to be relevant for the cosmological evolution today,
so as to resolve the coincidence problem, the scale for its effective 
mass must be 
set by the present value of the Hubble parameter \cite{wetcosmon}.
This means that the field is effectively massless at galactic length scales.
Its coupling to the dark matter particles results in a long-range 
force that can affect the details of structure formation 
\cite{peeblesfar,peeblessim}.

The presence of a long-range scalar interaction in addition to gravity
can modify the distribution of dark matter in galaxy halos. In particular,
it is not certain that the dark matter distribution can maintain a
form that results in approximately 
flat rotation curves for objects orbiting galaxies. 
Recent simulations of structure formation \cite{peeblessim} are not detailed
enough at the scales of interest in order to resolve this question. 

We consider an interaction between the scalar field and the
dark matter particles that can be modelled through 
a field-dependent particle mass. The action takes
the form
\begin{equation}
{\cal S}=\int d^4x \sqrt{-g}
\left(M^2 R -\frac{1}{2}g^{\mu\nu}
{\partial_\mu \phi}\,
{\partial_\nu \phi}
-U(\phi) \right)
-\sum_i \int m(\phi(x_i))d\tau_i
+{\cal S}_b,
\label{one} \end{equation}
where $d\tau_i=\sqrt{
-g_{\mu\nu}(x_i)dx^\mu_idx^\nu_i}$ and
the second integral is taken over particle trajectories. The contribution
${\cal S}_b$ describes the standard baryonic component. 
Variation of the action with respect to $\phi$ results in the
equation of motion 
\be
\frac{1}{\sqrt{-g}}{\partial_\mu}
\left(\sqrt{-g}\,\,g^{\mu\nu}{\partial_\nu \phi}
\right)=
\frac{dU}{d\phi}-\frac{d\ln m(\phi(x))}{d\phi}\,\, T^\mu_{~\mu},
\label{two} \ee
where the energy-momentum tensor associated with the gas of particles
is 
\be
T^{\mu\nu}=\frac{1}{\sqrt{-g}} 
\sum_i \int d\tau_i \,\, m(\phi(x_i))\,\, 
\frac{dx_i^\mu}{d \tau_i}\frac{dx_i^\nu}{d \tau_i}
\delta^{(4)}(x-x_i).
\label{three} \ee

We are interested in static spherically symmetric configurations,
with the scalar field varying slowly with the radial distance $r$. 
Our treatment is relevant up to 
a distance $r_1 \sim 100$ kpc beyond which the dark matter becomes very
dilute. For $r \gta r_1$ 
we expect 
that $\phi$ quickly becomes constant with a value close 
to $\phi(r_1)\equiv \phi_1$. 
This is the value that
drives the present cosmological expansion. Here we assume that the
cosmological evolution of 
$\phi_1$ is negligible for the time scales of interest, so that
the asymptotic configuration is static to a good approximation.

We approximate:
$m(\phi)\simeq m(\phi_0)+[dm(\phi_0)/d\phi]\, \dphi
\equiv m_0+m'_0 \, \dphi$, with $\phi_0$ the value of the field at 
the center of the galaxy ($r=0$). We work within the
leading order in $\dphi$ and assume that $m'/m\simeq m'_0/m_0$ 
for all $r$. Also
$dU/d\phi$ can be approximated by a constant between $r=0$ and $r=\infty$.
For the scalar field to provide a resolution of the coincidence problem,
the two terms in the r.h.s. of eq. (\ref{two}) must be of 
similar magnitude in the cosmological solution.
This means that 
$dU/d\phi$ must be comparable to $(m'_0/m_0)\rho_\infty$.
We expect $\rho_\infty$ to be a fraction of the critical density, i.e. 
$\rho_\infty \sim 3$ keV/cm$^3$. 
On the other hand, the energy density 
in the central region of the static
solution ($r \lta 100$ kpc) is that of the galaxy halo 
($\sim 0.4$ GeV/cm$^3$ for our neighborhood of the Milky Way).
This makes $dU/d\phi$ negligible in the r.h.s. of
eq. (\ref{two}) for a static configuration. The potential is
expected to become important only for $r \to \infty$, where the
static solution must be replaced by the cosmological one.
Similar arguments indicate that we
can neglect $U$ relative to $\rho$. Also the scalar field must be effectively
massless at the galactic scale. 
For these reasons we expect that the form of the 
potential plays a negligible role at the galactic level. Our analysis
can be carried out with $U=0$ and is model independent.

We treat 
the dark matter as a weakly interacting, dilute gas.
We are motivated by the phenomenological 
success of the isothermal sphere \cite{peeblesb} in 
describing the flat part of the rotation curves. 
We do not address the question of the density profile in the inner part of
the galaxies ($r \lta 5$ kpc). 
We approximate the energy-momentum tensor of the dark matter
as $T^\mu_{~\nu}={\rm diag} (-\rho,p,p,p)$ with
$p(r)=\rho(r)\, \langle \vrm_d^2 \rangle
=m(\phi(r))\, n(r)\, \langle \vrm_d^2 \rangle$. The 
dispersion 
%of each component 
of the dark matter velocity is
assumed to be constant and small: $\langle \vrm_d^2 \rangle \ll 1$. 
The gravitational field is considered in the 
Newtonian approximation: $g_{00}\simeq 1+2\Phi$, with
$\Phi = {\cal O}\left( m'_0\dphi/m_0\right)$. 
In the weak field limit and for
$p \ll \rho$, the conservation of the energy-momentum tensor gives 
\be
p'=-\rho\, \Phi'-\rho \, \frac{m'_0}{m_0} (\dphi)',
\label{extra} \ee
with the prime on $p$, $\Phi$, $\dphi$ denoting a derivative with 
respect to $r$. Integration of this equation gives
$n\simeq n_0
\exp \left(-\Phi/\langle \vrm_d^2 \rangle 
-(m'_0/m_0)\dphi/\langle \vrm_d^2 \rangle \right).$
%We can define an effective temperature $T_0$
%through the relation:
%$\langle \vrm_d^2 \rangle=T_0/m_0$. 
In order to make the picture more complete we also
allow for a pressureless baryonic component in the core. We assume that its
energy density has the phenomenological profile
$\rho_b(r)=\rho_B\, f \left( r/r_c \right)$, with $f(x)$ a
decreasing function of $x$.
We do not discuss the physics that leads to such a profile. 
We include the baryonic contribution only in order to 
estimate its effect on the dark matter distribution in the
outer regions of the galaxy.

With the above assumptions we obtain the equations of motion 
\be 
\Phi''+ \frac{2}{r}\Phi'=\frac{1}{4M^2} \, \rho_0 
\exp \left(-\alpha \Phi-\at \dphi \right)
+\frac{1}{4M^2} \, \rho_B \,\,
f \left(\frac{r}{r_c} \right),
\label{five} \ee
and 
\be
(\dphi)''+\frac{2}{r}(\dphi)'=\frac{m'_0}{m_0}\, \rho_0 
\exp \left(-\alpha \Phi-\at \dphi \right),
\label{six} \ee
where $M=(16\pi G_N)^{-1/2}$ is the reduced Planck mass, 
$\rho_0=m_0n_0$ the energy density of dark
matter at $r=0$, $\alpha
%=m_0/T_0
=1/\langle \vrm_d^2 \rangle$, 
and $\at=m'_0/(m_0 \langle \vrm_d^2 \rangle)$.
We emphasize that, even though $|\Phi| \ll 1$, the 
combination $\Phi/ \langle \vrm_d^2 \rangle$, 
that appears in the exponent in the
expression for the number density $n$, can be large.
Similarly, the expansion of the mass around the value 
$m_0=m(\phi_0)$ assumes the smallness of the dimensionless parameter 
$|m'_0 \dphi/m_0|$. However, the combination 
$\at \dphi=(m'_0 \dphi/m_0)/\langle \vrm_d^2 \rangle$, 
that appears in the exponent, can be
large.

A linear combination of eqs. (\ref{five}), (\ref{six}) gives
\be
\frac{d^2u}{dz^2}+\frac{2}{z}\frac{du}{dz}+\exp u
+\frac{R}{1+\kappa^2}\,\, f\left( \frac{z}{z_c} \right)=0,
\label{seven} \ee
where
$u=-\alpha \Phi-\at \dphi$, $z=\beta r$, $z_c=\beta r_c$,
$R={\rho_B}/{\rho_0}$, 
and $\beta^2=(1+\kappa^2)\alpha {\rho_0}/{4M^2}$. 
The parameter 
$\kappa^2=4M^2\left( {m'_0}/{m_0}\right)^2$
determines the strength of the new interaction relative to gravity. 
For large $z$ all regular solutions approach the
form $\exp u= 2/z^2$.
Another linear combination of eqs. (\ref{five}), (\ref{six}) gives 
\be 
\frac{d^2v}{dz^2}+\frac{2}{z}\frac{dv}{dz}
+\frac{\kappa^2R}{1+\kappa^2} \,\, f\left( \frac{z}{z_c}\right)=0,
\label{eight} \ee 
with 
$v=-\kappa^2 \alpha \Phi+\at\dphi.$
For large $z$ the solution of this equation is 
$v=c_0+{c_1}/{z}.$
Combining the two solutions gives the leading result for large $r$
\be
\Phi'=
\frac{2\langle \vrm_d^2 \rangle}{1+\kappa^2}
\frac{1}{r}.
\label{nine} \ee
If the new interaction is universal for ordinary and dark matter, the
experimental constraints impose $\kappa^2 \ll 1$. In this case, it is
reasonable to expect a negligible effect in the distribution of
matter in galaxy halos. 
However, if $\phi$ interacts only with dark matter, as we assume here,
this bound can 
be relaxed significantly.

According to eq. (\ref{nine}), 
a massive particle in orbit around the galaxy, at a large distance $r$ from
its center, has a velocity 
$\vrm_c^2=2\langle \vrm^2_{d} \rangle/(1+\kappa^2).$
We can use this expression in order to fix 
$\langle \vrm^2_d \rangle$ for
a given value of $\kappa$. 
The approximately 
flat rotation curves are a persistent feature 
even in the case that the dark matter is coupled to a scalar
field through
its mass. The effect of the new scalar interaction is encoded in the factor
$\kappa^2$. When this is small, the velocity of
an object orbiting the galaxy is of the order of the square root of
the dispersion of the 
dark matter velocity. If $\kappa^2$ is large, the 
rotation velocity can become much smaller than the typical 
dark matter velocity.

%The dimensionful quantity $\beta$
%can be written as
%$\beta^2={\rho_0}/(2M^2 \vrm^2_c).$ 
%In this way it is determined only through the observed rotation velocity 
%$\vrm_c$ and the density of dark matter at the center of the galaxy.
%The explicit strong dependence on $\kappa$ disappears. 
%This results in
%a profile for the rotation curves that is only weakly dependent on $\kappa$.
The Newtonian potential can be expressed as
$\Phi'=-2\left(u'+v'\right)/\vrm^2_c$.
This allows us to write the rotation velocity as
\be
\left(\frac{\vrm}{\vrm_c}\right)^2=\frac{r\Phi'}{\vrm^2_c}=-\frac{z}{2}
\left(\frac{du}{dz}+\frac{dv}{dz} \right).
\label{ten} \ee
We can also relate the 
parameters $z$ and $r$ through the expression
\be
z=\beta r\simeq\left(
\frac{\rho_0}{0.2\,M_\odot/{\rm pc}^3}
\right)^{1/2} \,
\frac{150~{\rm km/s}}{\vrm_c} \,\, \frac{r}{{\rm kpc}}.
\label{eleven} \ee

\begin{figure}[t]
 %\vspace{1.cm}
 \centerline{\epsfig{figure=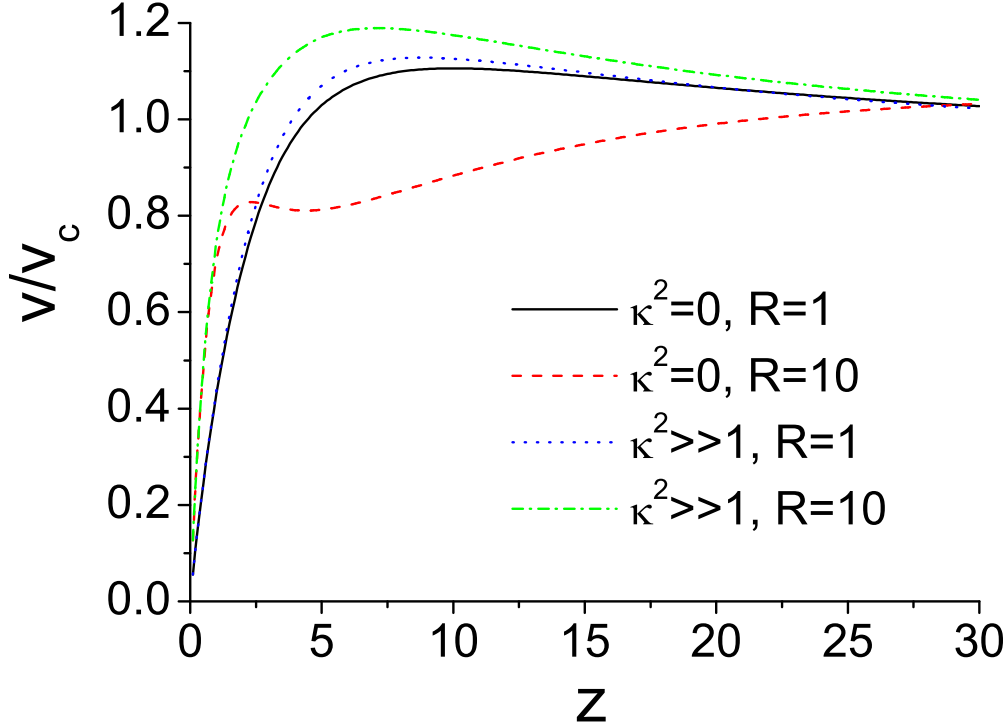,width=15cm,angle=0}}
 \caption{\it
The shape of the rotation curves for various values of $\kappa$ and $R$.
}
 \label{fig2}
 \end{figure}

In order to study the rotation velocity profile
more quantitatively we integrate
eqs. (\ref{seven}), (\ref{eight}) in the interval $0 \leq z \leq  30$.
We normalize the 
Newtonian potential $\Phi$ so that $\Phi(0)=0$. Also, according to our 
definition, $\dphi(r_0)=0$. These relations imply $u(0)=v(0)=0$.
The initial conditions $du(0)/dz=dv(0)/dz=0$
guarantee that 
the solution is regular at $r=0$.
In fig. 1 we present the solution of eqs. (\ref{seven}), (\ref{eight})
for various values of $\kappa$ and $R$. 
In all cases we use $f(x)=\exp(-x)$ and $z_c=0.5$.
We observe that the 
solutions do not differ substantially. 
Approximately flat rotation curves are expected for 
$z \gta 5$, consistently with observations \cite{peeblesb}. 
This behaviour persists for other forms of the profile of baryonic
matter, as long as dark matter dominates over
baryonic matter for $z\gta$ 3. The flat part of the rotation curves
is largely insensitive to the structure of the core.
The departure from spherical symmetry for the distribution of matter in
the core of a spiral galaxy is also expected to be a minor effect. 

Independently of the value of $\kappa^2$, the interaction of dark matter 
with the scalar field associated with dark energy does not destroy the
approximately 
flat profile of the rotation curves. Other considerations, however, could
constrain the coupling 
$\kappa^2=4M^2\left( {m'_0}/{m_0}\right)^2$. 
The dispersion of the dark matter velocity is 
$\langle \vrm^2_{d} \rangle=(1+\kappa^2) \vrm_c^2/2$. 
For a value of $\vrm_c$ deduced from observations, 
$\langle \vrm^2_{d} \rangle$ increases with $\kappa$. For sufficiently
large $\kappa$, it seems possible that $\vrm_c$
may exceed the escape velocity from the galaxy.
It turns out, however, that this is not the case. Outside the core of the 
galaxy and for
$r\lta r_1$, the binding potential for
a dark matter particle is $\Phi+(\at/\alpha)\dphi$. For large $r$, 
eq. (\ref{eight}) implies that $v=-\kappa^2\alpha\Phi+\at \dphi=$ constant. The
binding potential becomes $(1+\kappa^2)\Phi=(1+\kappa^2)\vrm^2_c\ln(r/r_1)$, 
where we have omitted an overall constant. For a particle
at a distance $r_*$ from the center of the galaxy, 
the escape velocity becomes
$\vrm^2_{esc}=2(1+\kappa^2)[\ln(r_1/r_*)+1]$. The value of $\vrm_{esc}$
is larger 
than the standard one \cite{spergel}
by a factor $(1+\kappa^2)$, 
so that $\langle \vrm^2_{d} \rangle$ remains substantially
smaller than $\vrm^2_{esc}$ for $r_* \ll r_1$. 
A particle that does not interact with the
scalar field is bound only by the the potential $\Phi$. However, the scale of
its velocity is set by $\vrm_c$, so that again it cannot escape.

Our discussion supports the conclusion that the presence in galaxies 
of a scalar field related to dark energy does not have any disastrous 
phenomenological consequences. 
On the contrary, this scenario may be much more interesting for dark 
matter searches than the conventional one. Depending on the value of $\kappa$,
the typical velocity of dark matter particles can exceed significantly the
observed rotation velocity ($\sim 220$ km/s for the Milky Way). 
As the density profile is not significantly modified, 
the estimated 
local energy density of dark matter remains the same 
as in the case with $\kappa^2=0$.
It is $\sim 0.4$ GeV/cm$^3$ for our neighborhood of the Milky Way. 
%(In order 
%for the solution of eq. (\ref{seven}) to reproduce this value, an appropriate
%adjustment of $R=\rho_B/\rho_0$ and $r_c$ may be necessary.) 
As a result the flux of dark matter particles towards a terrestrial
detector is larger roughly by a factor $(1+\kappa^2)^{1/2}$ relative to the 
$\kappa^2=0$ case. A detailed study must take into 
account the motion of the Earth through our galaxy. However, the Earth
velocity is of the order of $\vrm_c$, and gives only a modest correction.
The cross section for the elastic scattering of 
halo particles by target nuclei through weak or strong interactions
is largely independent of the particle velocity \cite{witten}.
The leading effect of a non-zero value of $\kappa$
is that the counting rates, that are
proportional to the velocity, are increased
by the factor $(1+\kappa^2)^{1/2}$. 
This makes the dark matter easier to detect. 
Existing bounds on dark matter properties from direct searches
can be extended to include the case of non-zero $\kappa$.
The bound on the cross section for the interaction  
of dark matter with the material of the detector must be strengthened by 
$(1+\kappa^2)^{1/2}$.

The allowed range of $\kappa$ is limited by the observable implications
of the model that describes the dark sector. Strong constraints arise
from the measured fluctuations of the cosmic microwave background. The 
dependence of the mass of 
dark matter particles on an evolving scalar field 
during the cosmological evolution since the 
decoupling is reflected in the microwave background. The magnitude of the
effect is strongly model dependent. In the models of ref. 
\cite{amendola,mainini}, which assume an exponential dependence of the 
dark matter mass on the field, the observations result in
the constraint $\kappa^2 \lta 0.01$. 
In the model of refs. \cite{peeblesfar,peeblessim}
the dark matter mass is a linear function of the field. The induced
scalar interaction among dark matter particles is screened by an additional
relativistic dark matter species. As a result, 
the model is viable even for couplings 
$\kappa^2 \simeq 1$. 
A similar mechanism is employed in ref. \cite{massimo} in
order to satisfy the observational constraints. In this model
the interaction between dark matter and dark energy 
becomes important only during the recent evolution
of the Universe.

It is reasonable to expect that 
the resolution of the coincidence problem through an interaction
between dark matter and dark energy will 
have to rely on a coupling not significantly
weaker that gravity. It seems unlikely that a coupling $\kappa^2\ll 1$ 
can lead to a cosmological evolution drastically different from 
that in the decoupled case. We expect 
the effects described in this letter to be
natural consequences of any viable model that achieves the resolution of the
coincidence problem through the coupling of dark matter to dark energy.

\vspace {0.5cm}
\noindent{\bf Acknowledgments}\\
\noindent 
I would like to thank C. Wetterich for many useful discussions. 
This work was 
%partially 
supported through 
%the RTN contract MRTN--CT--2004--503369 of the European Union, and 
the research program ``Pythagoras II'' (grant 70-03-7992) 
of the 
Ministry of National Education, partially funded by the European
Union.
%and ``Kapodistrias'' of the University of Athens (grant 70/4/6469). 

\end{document}